\documentclass[%
 superscriptaddress,
 reprint,
 amsmath,
 amssymb,
 prl,
 showpacs,
 floatfix,
 nofootinbib,
]{revtex4-1}

\usepackage{graphicx}
\usepackage{dcolumn}
\usepackage{bm}
\usepackage{hyperref}
\usepackage{color}
\usepackage{braket}
\usepackage{url}
\usepackage{xcolor}
\usepackage{soul}
\usepackage{tabularx}
\usepackage{multirow}
\usepackage{moresize}

\begin{document}

\title{Te Vacancy-Driven Anomalous Transport in ZrTe$_{5}$ and HfTe$_{5}$}

\author{Elizabeth A. Peterson} \affiliation{Theoretical Division, Los Alamos National Laboratory, Los Alamos, NM 87545, USA}
\author{Christopher Lane} \affiliation{Theoretical Division, Los Alamos National Laboratory, Los Alamos, NM 87545, USA}
\author{Jian-Xin Zhu} 
\affiliation{Theoretical Division, Los Alamos National Laboratory, Los Alamos, NM 87545, USA}
\affiliation{Center for Integrated Nanotechnologies, Los Alamos National Laboratory, Los Alamos, NM 87545, USA}

\begin{abstract}
The strongly sample-dependent anomalous transport properties observed in the layered Dirac materials ZrTe$_{5}$ and HfTe$_{5}$ are known to strongly correlate with the presence of Te vacancies. One particular phenomenon, a negative longitudinal magnetoresistance (NLMR), has been widely speculated to be a signature of broken chiral symmetry, which serves as a hallmark of chiral anomaly in these materials. However, the role of electronic structure in the microscopic mechanism behind the transport properties of these materials, most importantly their sample dependence, is poorly understood. This prompts the question as to whether the NLMR is a genuine signature of the chiral anomaly or if the transport properties in ZrTe$_{5}$ and HfTe$_{5}$ are merely artifacts of other factors, like structural disorder. In this work, the effect of Te vacancies on the electronic structure of ZrTe$_{5}$ and HfTe$_{5}$ is investigated via first-principles calculations to garner insight into how they may modulate the transport properties of these materials. Te vacancies serve two purposes: modification of the cell volume via effective compressive strain and also production of local changes to the electronic structure. The reorganization of the electronic structure near the Fermi energy indicates that Te vacancies can rationalize conflicting reports in spectroscopic and transport measurements that have remained elusive in prior first-principles studies. These results show that Te vacancies contribute, at least in part, to the anomalous transport properties of ZrTe$_{5}$ and HfTe$_{5}$ but, critically, do not eliminate the possibility of a genuine manifestation of the chiral anomaly in these materials.
\end{abstract}

\maketitle

\section{\label{sec:intro}Introduction}

Theoretical predictions of quantum anomalies, symmetry breaking that occurs when moving from classical field theories to quantum field theories, originated in the fields of particle and cosmological physics ~\cite{Adler1969, Bell1969, Kharzeev2006, Kharzeev2014}. As experimental verification of predicted quantum anomalies at the relevant length and energy scales of particle and cosmological physics is generally challenging or simply intractable, the possibility of observing quantum anomalies in experimentally accessible condensed matter systems has engendered a great deal of excitement. In the case of three-dimensional (3D) Dirac and Weyl semimetals, a chiral anomaly is characterized by an imbalance in right- and left-handed chiral fermions in the presence of an applied magnetic field. The resulting chiral current may be induced by application of parallel electric and magnetic fields producing an anomalous contribution to the conductivity ~\cite{Nielsen1983, Fukushima2008, Son2013, Kharzeev2014, Huang2015, Xiong2015, Arnold2016, Zhang2016, Li2016, Chi2017, Nandy2017, Nag2020, Nandy2021, Xie2022}. This manifests as a negative longitudinal magnetoresistance (NLMR), an experimental signature of the chiral anomaly. In recent years the layered 3D Dirac materials ZrTe$_{5}$ and HfTe$_{5}$ have garnered attention as material platforms that potentially host a chiral anomaly ~\cite{Li2016, Zheng2016, Chi2017}. These materials exhibit a number of anomalous transport properties, including a Lifshitz transition as a function of temperature and a negative longitudinal magnetoresistance ~\cite{Chen2015B, Li2016, Liu2016, Zheng2016, Chi2017, Lv2018, Zhang2020, Xie2022, Kovacs2023, Jo2023}. This NLMR is suggested to be evidence of the chiral anomaly in these materials ~\cite{Li2016, Zheng2016, Chi2017}. However, NLMR is not sufficient to prove the presence of a chiral anomaly, as it may arise from alternative sources such as current jetting or defects ~\cite{Arnold2016, Reis2016, Li2017}. Lacking a robust microscopic description of the source of the anomalous transport properties in ZrTe$_{5}$ and HfTe$_{5}$, it is as yet unclear if these materials exhibit a genuine chiral anomaly. Further, there is widespread controversy in the characterization of the topological nature of these materials; different experiments report different topological phases, including topological insulating and Dirac semimetallic states ~\cite{Weng2014, Chen2015, Chen2015B, Liu2016, Li2016, Zheng2016, XiangBing2016, Wu2016, Moreschini2016, Manzoni2016, Tajkov2022, Kovacs2023}.

The sample dependence of the topological ~\cite{Manzoni2016, Fan2017, Zhang2017, Monserrt2019, Jo2023} and transport ~\cite{Shahi2018, Lv2018, Jiang2023, Jo2023} properties of ZrTe$_{5}$ and HfTe$_{5}$ are well documented. Crystal growth by different methods, such as flux growth and chemical vapor transport (CVT), or different preparation processes utilizing the same method, may produce samples with different lattice constants and different (off-)stoichiometries. One suspected culprit for this variability is the production of different types and concentrations of point defects, including Te vacancies and Te interstitials. The variability in topological and transport properties is further surmised to depend on sample unit cell volume.

Topological phase transitions from a strong topological insulator, to a Dirac semimetal, to a weak topological insulator are theoretically predicted to occur with increasing volume ~\cite{Manzoni2016, Zhang2017, Fan2017, Monserrt2019, Tajkov2022, Jo2023}. Recent experimental and theoretical results using angle-resolved photoemission spectroscopy (ARPES) and first-principles calculations rationalize the observed sample dependence of topological phases by reporting that Te vacancies are a source of chemical pressure, or internal strain, altering the volume of ZrTe$_{5}$ and HfTe$_{5}$~\cite{Jo2023}.

The Lifshitz transition (experimentally characterized by a change from p-type to n-type carriers and from metallic to semiconducting to metallic character with decreasing temperature as measured through Hall and longitudinal conductivity, respectively) that is widely featured in the literature on ZrTe$_{5}$ and HfTe$_{5}$ does not occur in all samples ~\cite{Shahi2018, Lv2018, Jiang2023, Jo2023}. In fact, Te deficiency appears to be critical to the observation of the Lifshitz transition at all. Experimental studies of stoichiometric and Te-deficient ZrTe$_{5}$ and HfTe$_{5}$ consistently show that the anomalous longitudinal resistivity peak only occurs in Te-deficient samples; samples that are highly stoichiometric exhibit a monotonic insulator-like increasing resistivity down to low temperatures ~\cite{Shahi2018, Lv2018}. Concomitantly, the change in the sign of the Hall resistivity was measured to correlate with Te-deficiency as well ~\cite{Shahi2018}.

Outstanding discrepancies between low temperature spectroscopic and transport measurements of ZrTe$_{5}$ and HfTe$_{5}$ suggest that the significance of Te vacancies evinces richer physics including, but not limited to, volume effects. Below the Lifshitz transition temperature, Hall conductivity measurements indicate that the primary carriers in these materials are electrons, suggesting the conduction band must be populated ~\cite{Liu2016,Chi2017,Shahi2018,Lv2018,Kovacs2023}. While most reported ARPES measurements show that as temperature decreases the Fermi energy shifts downward into the valence band at the $\Gamma$ point, suggesting the primary carriers are holes ~\cite{Li2016, Wu2016, Moreschini2016}; one ARPES study observed that as temperature decreases the Fermi energy shifts upward into the conduction band in the neighborhood of the $\Gamma$ point suggesting the primary carriers are electrons, consistent with Hall conductivity measurements ~\cite{Zhang2017}. This study went on to measure the full Brillouin zone (BZ), identifying electron pockets far from the BZ center. Electronic structure calculations on pristine ZrTe$_{5}$ and HfTe$_{5}$ do not predict any portion of the conduction band dipping below the Fermi energy, an apparent contradiction. Moreover, ARPES measurements generally suggest the Dirac point is centered at the $\Gamma$ point while electronic structure calculations of the pristine materials generally place the Dirac point along the $\Gamma$-Y high-symmetry line (or $\Gamma$-Z depending on convention) ~\cite{Li2016, Wu2016, Moreschini2016, Zhang2017, Shahi2018, Jo2023, Kovacs2023}.

The observation that the Lifshitz transition and NLMR occur only when Te vacancies are present strongly indicates that Te vacancies serve as more than just a source of effective strain in altering the electronic structure and transport properties of ZrTe$_{5}$ and HfTe$_{5}$. However, a comprehensive picture of the effect of Te vacancies on the electronic structure of these materials remains to be fully demonstrated. This microscopic picture is critical to disentangling and addressing two questions: (a) to what extent can the conflicting transport and spectroscopic measurements be rationalized by Te vacancies? and (b) is the NLMR observed in these materials actually a signature of a chiral anomaly or merely an artifact of structural disorder? In this work, the first of these questions is addressed and a preliminary assessment of the second is put forward. Here, first-principles density functional theory (DFT) calculations of the electronic structure of ZrTe$_{5}$ and HfTe$_{5}$ in the presence of Te vacancies offer insight into the role that Te vacancies play in modulating the electronic structure of these materials. These calculations reveal that while Te vacancies do indeed serve as a source of chemical pressure, or effective strain, they also significantly modify the electronic structure of ZrTe$_{5}$ and HfTe$_{5}$. The resulting modifications are interpreted in light of the diversity of reported experimental measurements and offer a common thread rationalizing seemingly conflicting measurements. They also leave open the possibility that a chiral anomaly does indeed occur in these materials.

\section{\label{sec:results}Electronic Structure Calculations}

\begin{figure}
\includegraphics[width=0.9\linewidth]{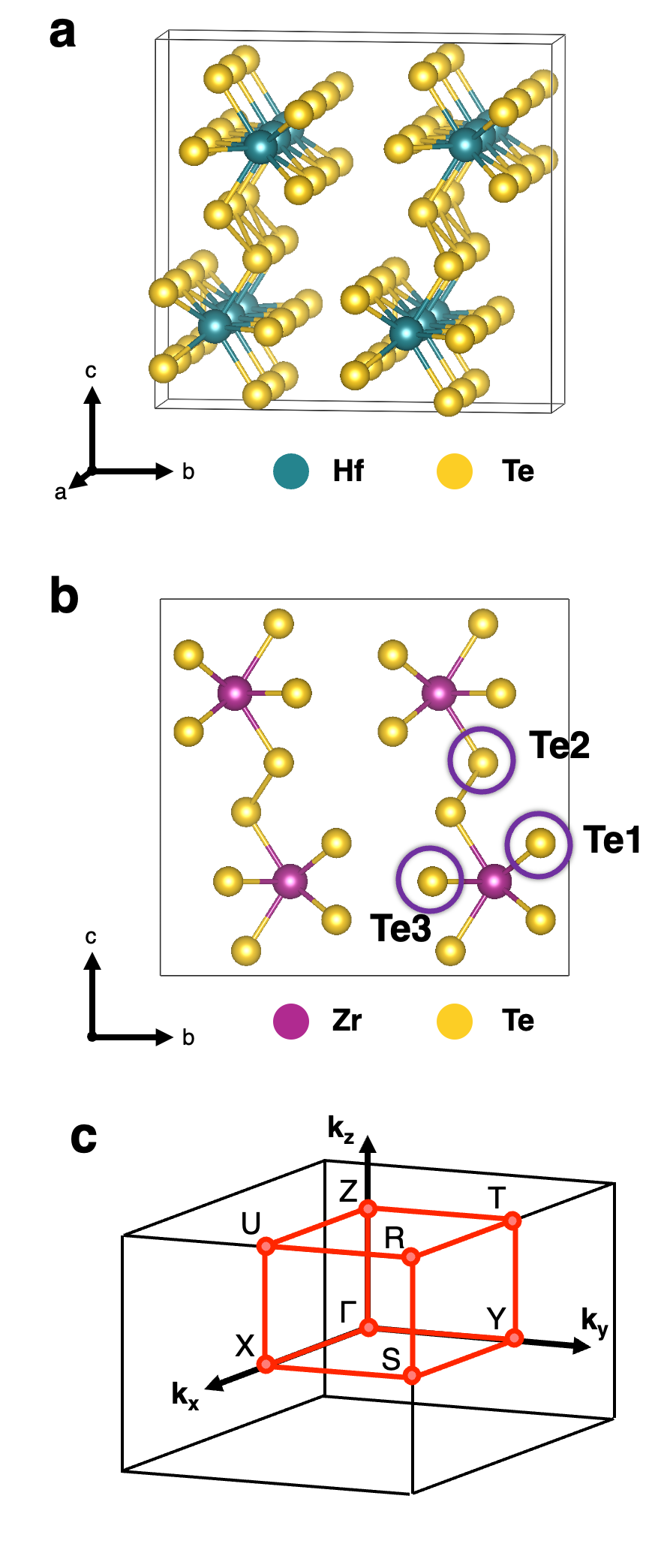}
\caption{\label{fig:crystal_structures}(a) The conventional crystal structure of HfTe$_{5}$ characterized by Hf ion (teal) polyhedron centers each coordinated to 8 Te ions (yellow). Layers of HfTe$_{8}$ polyhedra are formed by Hf-Te bonded chains along the $\mathbf{a}$ direction and Te-Te bonded chains along the $\mathbf{c}$ directions. Layers are stacked along the $\mathbf{b}$ direction held together by vdW dispersion forces. (b) The isostructural conventional crystal structure of ZrTe$_{5}$ with Zr ions in purple and Te ions in yellow. The three symmetrically distinct Te ion sites are indicated. (c) Orthorhombic Brillouin zone with high-symmetry points used in the band structure calculations shown in Figure 2-4 marked.}
\end{figure}

ZrTe$_{5}$ and HfTe$_{5}$ are layered materials with one-dimensional chains of HfTe$_{8}$ polyhedra along the $\mathbf{a}$ axis connected by Te-Te bonds along the $\mathbf{c}$ axis. These ac-planes are bound by vdW dispersion forces along the $\mathbf{b}$ axis, as shown in \textbf{Figure \ref{fig:crystal_structures}}(a). The lattice constants of ZrTe$_{5}$ and HfTe$_{5}$ are similar, as shown in Table \ref{table_latt}. They have three symmetrically distinct Te sites, as shown in Figure \ref{fig:crystal_structures}(b). Prior experiments suggest that Te vacancies form readily on sites 2 and 3 in ZrTe$_{5}$ ~\cite{Shahi2018}, however the calculated formation energies of Te vacancies on each site indicate that a vacancy on site 1 is the most favorable, in agreement with recent work ~\cite{Jo2023}. The DFT calculated formation energies for Te vacancies on each site are shown in Table \ref{table_zrte5_hfte5}. They are all relatively high compared to average thermal energy $k_{b}T$ ($\sim$ 26 meV) at room temperature, ranging from 77-98 meV/atom. Nonetheless, under high temperature or non-equilibrium synthesis conditions a finite population of Te vacancies would still be expected to form. Notably, the relative formation energy differences between different vacancy sites are on the order of 10-20 meV/atom, meaning that a mixture of vacancy types should be present even though site 1 is the most stable.

\begin{table}[h]
 \caption{Lattice constants of the experimental crystal structures of ZrTe$_{5}$ and HfTe$_{5}$ from the ICSD ~\cite{Fjellvag1986, Furuseth1973} and the fully geometrically relaxed crystal structures calculated using DFT with Grimme-D3 dispersion corrections.}
 \begin{center}
  \begin{tabular}{c  c  c  c  c}
    \hline
    & Source & a (\textrm{\AA}) & b (\textrm{\AA}) & c ({\AA}) \\
    \hline
    ZrTe$_{5}$ & ICSD & 3.987 & 14.530 & 13.724  \\
    & DFT & 4.026 & 14.793 & 13.636 \\
    HfTe$_{5}$ & ICSD & 3.968 & 14.455 & 13.691  \\
    & DFT & 4.000 & 14.711 & 13.596 \\
    \hline
  \end{tabular}
  \end{center}
  \label{table_latt}
\end{table}

\begin{table*}
 \caption{Defect formation energies and changes in the volume and lattice constants for ZrTe$_{5}$ and HfTe$_{5}$ with Te vacancies at each symmetrically distinct site. Changes in the volume and lattice constants are for the fully relaxed defect structure and reported relative to the volume and lattice constants of the fully relaxed pristine crystal structure.}
 \begin{center}
  \begin{tabular}{c c c c c c c}
    \hline
    & V$_{\textrm{Te}}$ Site & E$_{f}$ (meV/atom) & $\Delta$V ($\%$) & $\Delta$a ($\%$) & $\Delta$b ($\%$) & $\Delta$c ($\%$) \\
    \hline
    ZrTe$_{5}$ & 1 & 83.1 & -0.14 & -0.10 & 0.00 & -0.03  \\
    & 2 & 88.2 & -1.42 & -1.01 & -0.07 & -0.35 \\
    & 3 & 97.6 & -1.02 & -0.19 & 0.00 & -0.83  \\
    HfTe$_{5}$ & 1 & 77.0 & -0.08 & -0.12 & +0.06 & -0.02 \\
    & 2 & 83.5 & -1.42 & -1.09 & -0.03 & -0.31 \\
    & 3 & 94.8 & -1.10 & -0.27 & +0.11 & -0.93  \\
    \hline
  \end{tabular}
  \end{center}
  \label{table_zrte5_hfte5}
\end{table*}

\subsection{Band Structures}

\begin{figure*}
  \includegraphics[width=0.9\textwidth]{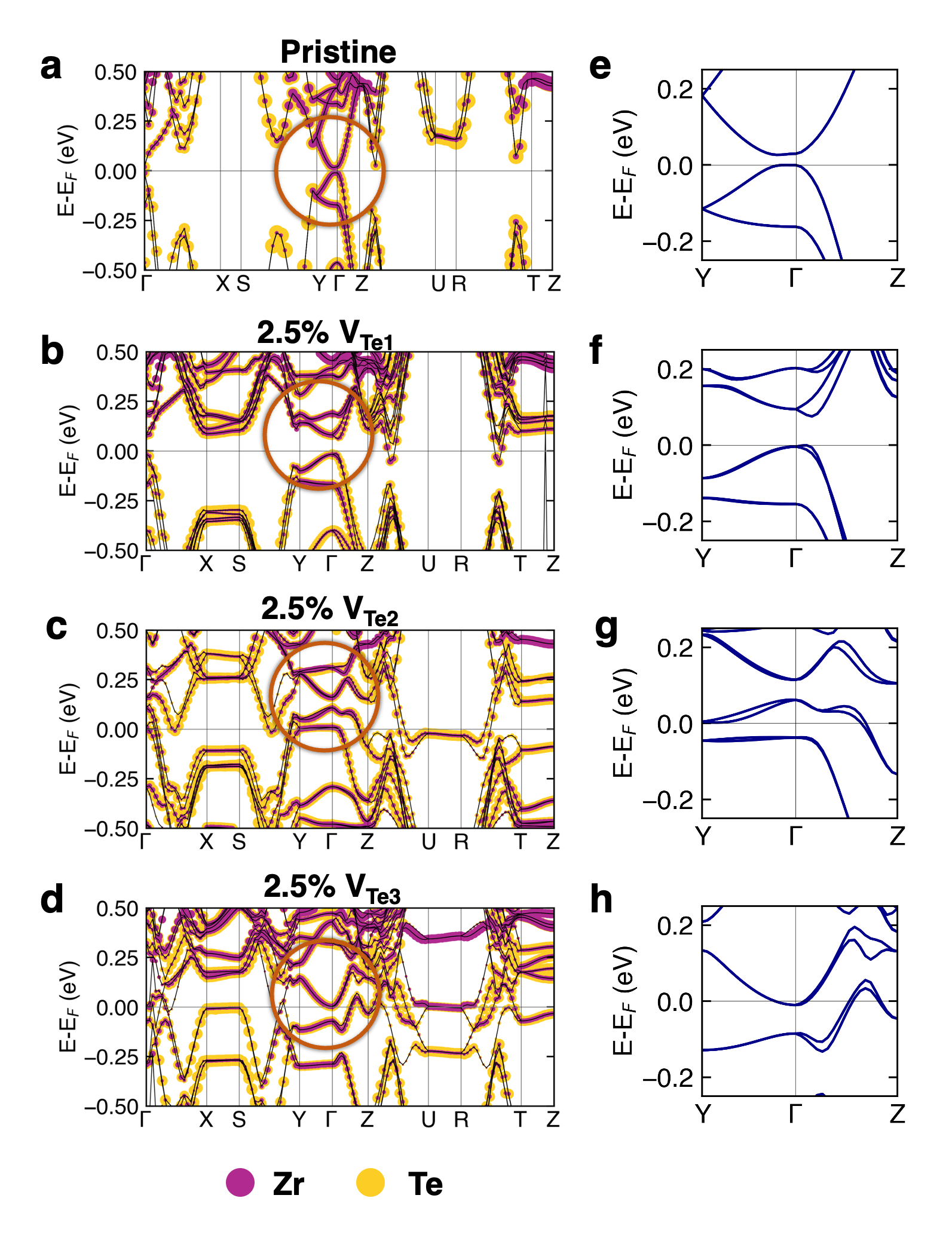}
  \caption{The DFT+SOC band structures near the Fermi energy of ZrTe$_{5}$ in the fully relaxed geometry for the pristine crystal structure (a,e) and the cases of a neutral Te vacancy in Te site 1 (b,f), Te site 2 (c,g), and Te site 3 (d,h). High symmetry lines in the entire orthorhombic Brillouin zone shown in Figure 1c are plotted in (a-d). 
  Zoomed-in views close to the $\Gamma$-point are plotted in (e-h).}
  \label{fig:zrte5_band_structures}
\end{figure*}

\begin{figure*}
  \includegraphics[width=0.9\textwidth]{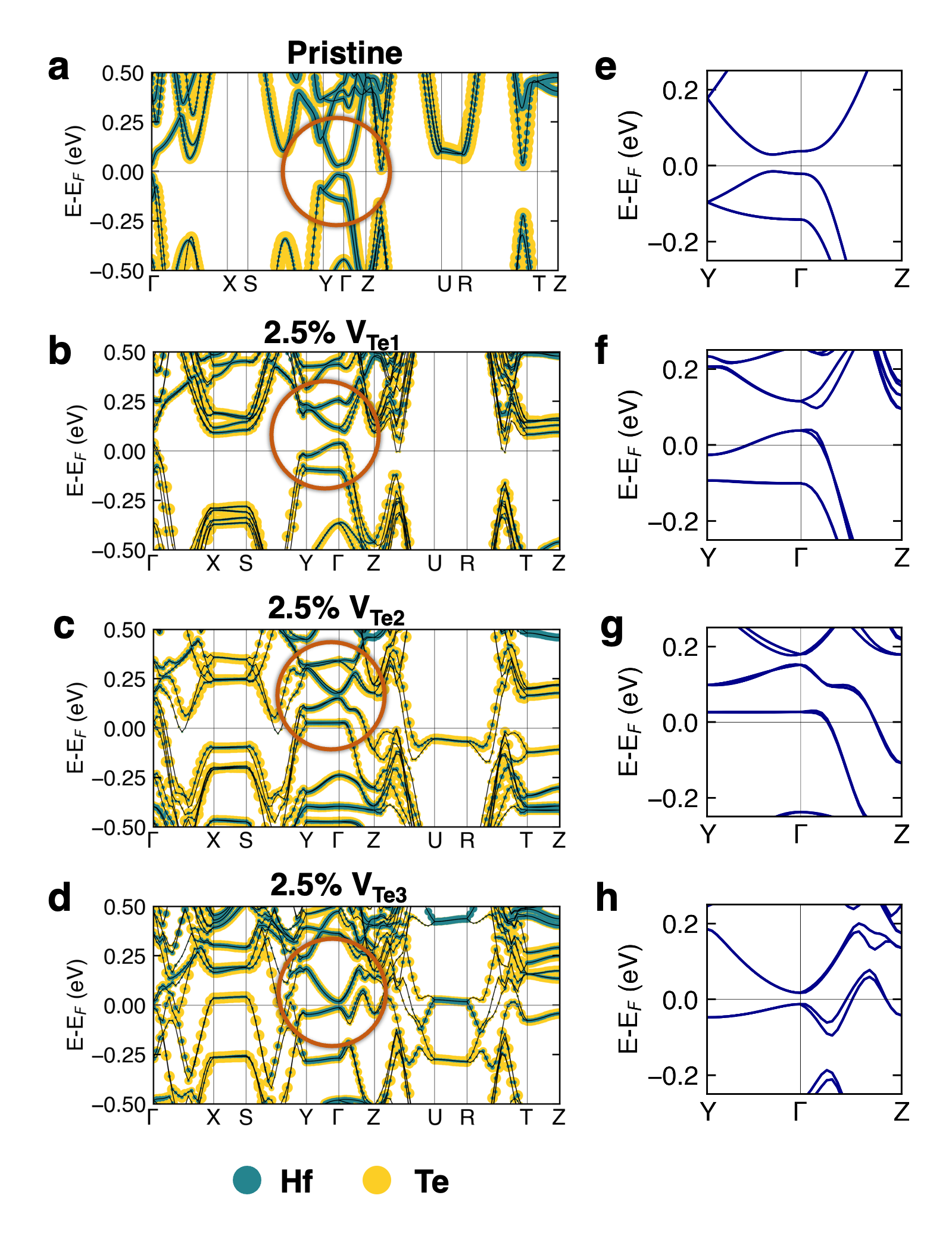}
  \caption{Same as Figure \ref{fig:zrte5_band_structures} except for HfTe$_{5}$. The DFT+SOC band structures near the Fermi energy of ZrTe$_{5}$ in the fully relaxed geometry for the pristine crystal structure (a,e) and the cases of a neutral Te vacancy in Te site 1 (b,f), Te site 2 (c,g), and Te site 3 (d,h). High symmetry lines in the entire orthorhombic Brillouin zone shown in Figure 1c are plotted in (a-d). Zoomed-in views close to the $\Gamma$-point are plotted in (e-h).}
  \label{fig:hfte5_band_structures}
\end{figure*}

The band structures of pristine ZrTe$_{5}$ and HfTe$_{5}$ (calculated with DFT at the GGA-PBE level of theory) are both characterized by a Dirac point at the Fermi energy that opens into a narrow band gap between the $\Gamma$ and Y high-symmetry points due to spin-orbit coupling (SOC) as shown in \textbf{Figure \ref{fig:zrte5_band_structures}}(a,e) and \textbf{Figure \ref{fig:hfte5_band_structures}}(a,e) (see Supporting Information \textbf{Figure S1} for band structures without SOC). The PBE+SOC calculated indirect band gaps of pristine ZrTe$_{5}$ and HfTe$_{5}$ are 27 meV and 24 meV respectively. When Te vacancies are introduced, the electronic structure calculated with SOC looks very similar for both ZrTe$_{5}$ and HfTe$_{5}$. This is likely a reflection of the very close similarity in lattice parameters between the two isostructural materials.

The PBE+SOC band structures for Te vacancies at site 1 (Figure \ref{fig:zrte5_band_structures}(b,f) and Figure \ref{fig:hfte5_band_structures}(b,f)) are quite similar to the pristine cases except with (i) the direct band gap at the Dirac point shifted from being along the $\Gamma$-Y high-symmetry line to being exactly at the $\Gamma$ point and (ii) a decrease of the conduction band (CB) energy below the Fermi level along the Z-U and R-T high-symmetry lines. In ZrTe$_{5}$, the Fermi energy still lies within the Dirac point gap while in HfTe$_{5}$ the Fermi energy cuts through the valence band (VB) near the Dirac point at $\Gamma$. Te vacancies at site 2 (Figure \ref{fig:zrte5_band_structures}(c,g) and Figure \ref{fig:hfte5_band_structures}(c,g)) shift the Fermi energy below the VB top at the Dirac point and also shift the CB below the Fermi energy in several more regions of the Brillouin zone (BZ). Te vacancies at site 3 (Figure \ref{fig:zrte5_band_structures}(d,h) and Figure \ref{fig:hfte5_band_structures}(d,h)) shift the CB and VB in opposite directions both towards the Fermi energy.

The electron and hole pockets introduced near the Fermi energy by each type of Te vacancy indicate that a combination of hole- and electron-like carriers should exist at low temperatures. Te vacancies at site 1 produce a strong excess of electron-like carriers, while vacancies at site 2 and 3 produce a more equal mix of hole- and electron-like carriers (see \textbf{Figure S3} in the Supporting Information for more details). As Te vacancies at site 1 are the most stable and should proliferate at the highest concentrations, these results support the observation of n-type conductivity at low temperatures.

Figure \ref{fig:zrte5_band_structures}(a-d) shows the band structures of pristine and Te-vacancy ZrTe$_{5}$ plotted with orbital contributions from each element indicated by purple and yellow dots for Zr and Te respectively. The band gap at the Dirac point, circled in orange, is consistently characterized by a mix of contributions from Zr and Te orbitals. The conduction bands that shift below the Fermi energy are predominantly of Te character. One notable exception is a flat band of Zr character that lies right at the Fermi energy along the U-R high-symmetry line for Te vacancies at site 3, which would be expected to have important implications for the transport properties of this case. Nearly identical observations can be made for the band structures of HfTe$_{5}$ as shown in \textbf{Figure \ref{fig:hfte5_band_structures}}(a-d).

In all cases, Te vacancies make the band structures of ZrTe$_{5}$ and HfTe$_{5}$ metallic, although least dramatically for Te vacancies at site 1, the most favorable Te vacancy. The volume reduction caused by Te vacancies at site 3 is the largest ($>1\%$), followed by site 2, while the volume reduction caused by Te vacancies at site 1 is much less dramatic ($<0.1\%$)(see Table \ref{table_zrte5_hfte5} for details). This explains the trend in metallicity across the band structures of each type of Te vacancy, with more effective compressive strain resulting in a more metallic band structure.

To ensure that the metallicity and reorganization of the bands at the Fermi level are not an artifact of limitations of the PBE+D3 method, the band structure of each vacancy site is calculated for HfTe$_{5}$ using lattice constants (a) fixed at the values for the pristine structures, allowing only internal coordinate relaxation, and (b) from a full geometry optimization. The band structures, reported in the Supporting Information \textbf{Figure S2}, show that volume effects only slightly alter the energies, at which the new conduction and valence band states appear near the Fermi energy, but do not qualitatively change the presence or dispersion of the new bands.

At the $\Gamma$ point, the direct band gap is larger in the presence of Te vacancies than it is for the pristine crystal structures. This is consistent with recent band structure calculations of HfTe$_{5}$ under compressive strain ~\cite{Jo2023} and supports the hypothesis that Te vacancies affect the electronic structure, at least in part, in a similar manner to an applied strain.

\subsection{Site Projected Densities of States}

\begin{figure*}
  \includegraphics[width=0.9\textwidth]{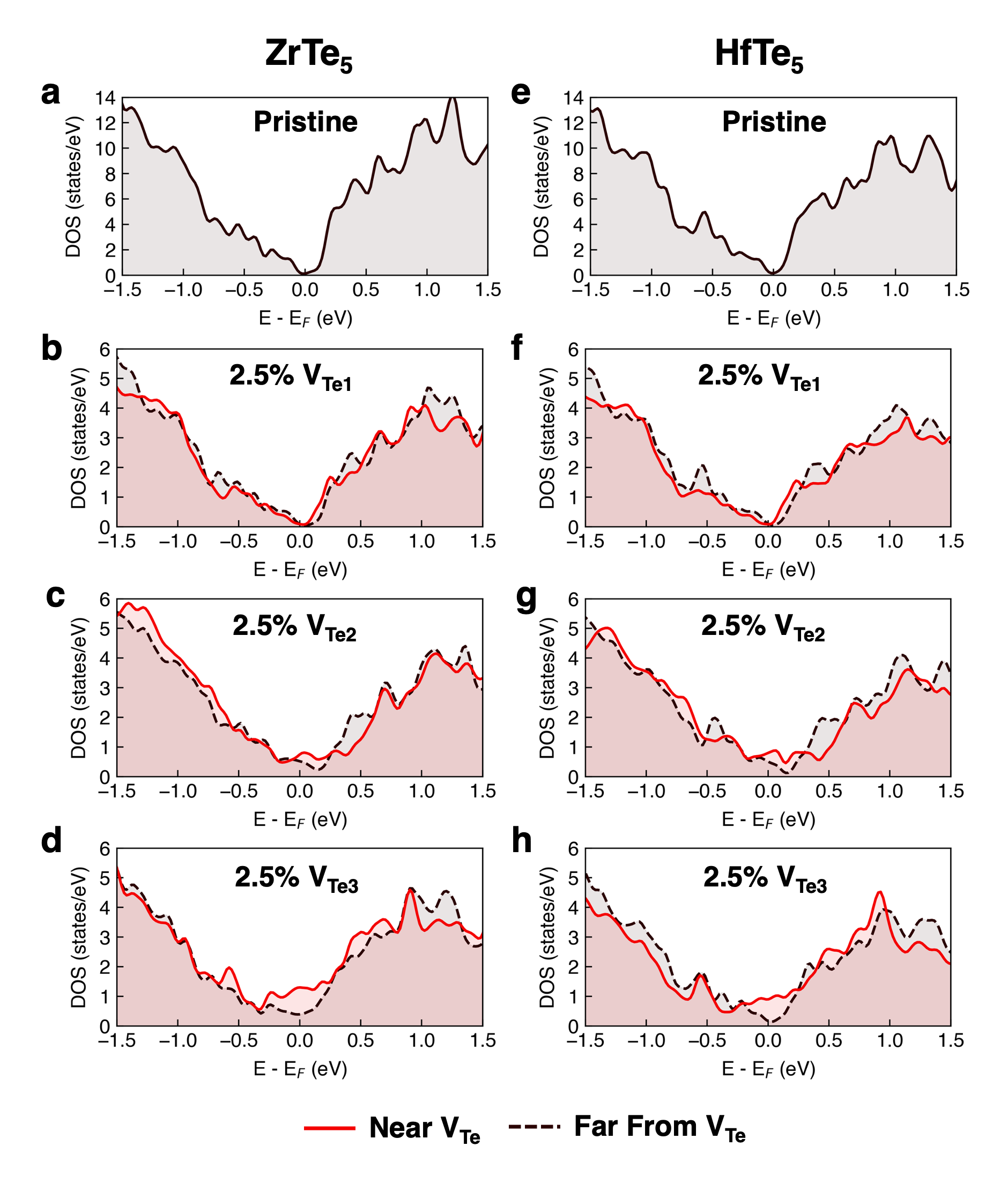}
  \caption{The DFT+SOC densities of states for ZrTe$_{5}$ and HfTe$_{5}$ in the fully relaxed geometries for the pristine crystal structure (a,e) and for a neutral Te vacancy on Te site 1 (b,f), Te site 2 (c,g), and Te site 3 (d,h). The projected density of states is plotted for orbitals on ions in the Zr (or Hf) polyhedron where the Te vacancy is located (Near V$_{Te}$, solid red line) and the Zr (or Hf) polyhedron farthest from the Te vacancy (Far from V$_{Te}$, black dashed line).}
  \label{fig:site_projected_dos}
\end{figure*}

To better understand the source of the energy shifts and reordering of the bands near the Fermi energy, the site-projected density of states is calculated for ions in regions near the Te vacancy sites and ions in regions far from the Te vacancy sites. This makes it possible to disentangle the distinct effects of volume compression and additional states contributed by the Te vacancies. The density of states far from the Te vacancy sites should reflect only the effects of compressive strain while the density of states close to the Te vacancy sites should reflect the combined effects of compression and any additional changes to the electronic structure generated by the presence of the Te vacancy. 

The local site projected density of states for the Zr and Te ions in the Zr polyhedron furthest away from the Te vacancy (Far from V$_{\textrm{Te}}$) strongly resembles the density of states of pristine ZrTe$_{5}$ (as shown in \textbf{Figure \ref{fig:site_projected_dos}}(a-d)), with the exception of a notable shift in the position of the band gap for Te vacancies on site 2. Conversely, the local site projected density of states for the ions in the Zr polyhedron where the Te vacancy is located (Near V$_{\textrm{Te}}$) indicates that orbitals near the Te vacancy are the primary source of in-gap states. A similar conclusion is drawn from the site projected densities of states both near and far from the Te vacancies for HfTe$_{5}$ (Figure \ref{fig:site_projected_dos}(e-h)). 

\section{\label{sec:discussion}Discussion}

The electronic structure calculations offer a robust avenue towards rationalizing the experimentally observed low temperature transport and spectroscopic properties of ZrTe$_{5}$ and HfTe$_{5}$, as  DFT calculations correspond to zero temperature. 

Near the $\Gamma$ point in the electronic structure, the Fermi energy either remains in the band gap at the Dirac point or cuts through the valence band (as in the case of Te vacancies at site 2). This is consistent with most of the experimental ARPES observations near the $\Gamma$ point ~\cite{Li2016, Wu2016, Moreschini2016}. However, when considering the entire Brillouin zone, it becomes clear that there is significant population of the conduction bands throughout regions of the BZ that are further away from $\Gamma$ when Te vacancies are present. This is consistent with more comprehensive ARPES measurements that observed electron pockets along the Y-S high-symmetry line ~\cite{Zhang2017}. 

Resistivity measurements at low temperature in Te deficient samples report metallic conduction with electron-like carriers. The low temperature metallicity is consistent with the existence of in-gap states in the vicinity of Te vacancies. In particular, Te vacancies at sites 2 and 3 can generate flat bands near the Fermi energy from conduction bands that have shifted to lower energy, such as in Figure \ref{fig:zrte5_band_structures}(c,d) and Figure \ref{fig:hfte5_band_structures}(c,d). From the band structure calculations, it becomes clear that a combination of populations of electron-like and hole-like carriers is to be expected at low temperature. However, flat bands associated with valence band states appear as well. Te vacancies at site 1, which are the most stable and expected to be the most prolific, produce a strong excess of electron-like carriers, consistent with Hall resistivity measurements that consistently observe n-type conductivity at low temperatures.

Combined, these calculations offer an explanation for the seemingly conflicting results of ARPES measurements and Hall conductivity measurements.

The production of in-gap states localized near anion vacancies is not a unique feature of Dirac materials. This is commonly observed, such as the production of in-gap flat bands spanning the entire Brillouin zone when oxygen vacancies are introduced into metal oxides ~\cite{Hinuma2018, Fernandez2020, Zhou2022}. The in-gap states produced by orbitals near the Te vacancies, however, are noteworthy in that they have a variety of dispersions near the Fermi energy, including linear crossings and flat bands along certain high symmetry lines. It is also important to note that they appear at points in the Brillouin zone far from the $\Gamma$ point. The narrow gap at the $\Gamma$ point that is the hallmark of massive Dirac materials is robust to the introduction of these point defects, as shown in Figure \ref{fig:zrte5_band_structures}(e-h) and Figure \ref{fig:hfte5_band_structures}(e-h). This robustness of the Dirac point against structural disorder is critical. It implies that the Dirac physics expected to enable a chiral anomaly could also be robust against structural disorder. 

\section{\label{sec:summary}Conclusions}

In this work, first-principles DFT electronic structure calculations of the layered Dirac materials ZrTe$_{5}$ and HfTe$_{5}$ with Te vacancies offer insight into the sample dependence of their experimentally observed spectroscopic and transport properties. Consistent with most low temperature ARPES measurements, Te vacancies are shown to promote both semiconducting or hole-doped behavior at the $\Gamma$ point. They further promote occupation of the conduction band by shifting conduction band states towards the Fermi energy which rationalizes the n-type conductivity at low temperatures consistently reported by Hall measurements. Different combinations of Te vacancies at different concentrations are expected to produce a spectrum of electronic structure reorganizations accounting for sample-dependent reports of gapped and metallic behavior. Te vacancies affect the electronic structure and corresponding transport properties of ZrTe$_{5}$ and HfTe$_{5}$ both as a general source of effective strain and, importantly, by introducing additional states near the Fermi energy localized in the vicinity of the Te vacancy sites. Critically, the existence of the gapped Dirac point in the DFT+SOC calculations is robust to the introduction of Te vacancies. The effect of Te vacancies on the topological character of the bands at the $\Gamma$ point is left to future work. However, these results suggest that the proposed mechanism of a chiral anomaly occurring in ZrTe$_{5}$ and HfTe$_{5}$ deriving from the topological nature of Dirac materials may be robust to structural disorder; moreover, this implies that the presence of defects does not eliminate the  possibility that the NLMR observed in ZrTe$_{5}$ and HfTe$_{5}$ may be a genuine signature of a quantum anomaly in experimentally accessible condensed matter systems.

\section{\label{sec:calc_details}Calculation Details}
First-principles calculations are performed using density functional theory (DFT) with a plane-wave basis and projector augmented wave (PAW) pseudopotentials ~\cite{Kresse1999} as implemented in the Vienna \textit{ab initio} simulation package (VASP) ~\cite{Kresse1996a,Kresse1996b}. Calculations are performed in the generalized gradient approximation (GGA) as implemented by Perdew, Burke, and Ernzerhof (PBE) ~\cite{Perdew1996} with additional vdW dispersion forces approximately accounted for via the Grimme-D3 method ~\cite{Grimme2010}. The crystal structures of bulk ZrTe$_{5}$ and HfTe$_{5}$ are relaxed using a 600 eV energy cutoff and $20\,\times8\,\times8$ $\Gamma$-centered k-mesh until forces are converged to $<$ 1 meV/\AA. This method results in good agreement of the $\mathrm{a}$ and $\mathrm{c}$ lattice constants (Table \ref{table_latt}). The out-of-plane $\mathrm{b}$ lattice constants are only slightly overestimated by 1.8$\%$ and 1.7$\%$ for ZrTe$_{5}$ and HfTe$_{5}$ respectively (see Table \ref{table_latt} for details). The band structure is calculated both with and without spin-orbit coupling (SOC) using a 500 eV energy cutoff. The density of states is calculated with SOC.

Te vacancy calculations are performed on $2\,\times1\,\times1$ supercells of bulk ZrTe$_{5}$ and HfTe$_{5}$ with a single Te ion removed, a 2.5$\%$ Te vacancy concentration. Vacancies are introduced at each of the three symmetrically distinct Te sites. Geometry optimization is performed both for fixed lattice parameters with only internal coordinates allowed to relax as well as full crystal structure relaxation. For the supercell calculations, an energy cutoff of 500 eV and a $10\,\times8\,\times8$ $\Gamma$-centered k-mesh are used. The band structure is calculated for the fully relaxed crystal structures, to capture volume effects, both with and without spin-orbit coupling (SOC) and the density of states is calculated with SOC. 

The defect formation energy is calculated using 

\begin{equation}
E_{f} = E_{tot}[\textrm{V}_{\textrm{Te}}] - E_{tot}[\textrm{pristine}] - \Sigma_{i}n_{i}\mu_{i}
\end{equation}

\noindent where $E_{tot}[\textrm{V}_{\textrm{Te}}]$ and $E_{tot}[\textrm{pristine}]$ are the DFT calculated total energies of the crystal structure with a Te vacancy and the pristine crystal structure. $n_{i}$ is the number of atoms of species $i$ added or removed (in this case one Te atom) and $\mu_{i}$ is the chemical potential of that species, here taken from a DFT calculation of bulk Te in the trigonal P3$_{1}$21 space group as tabulated in the ICSD ~\cite{Adenis1989}. The standard additional terms to correct for charge effects are neglected because only neutral Te vacancies are considered ~\cite{Vandewalle2004}.

\section{\label{sec:ackn}Acknowledgements}
This work was supported by the U.S. DOE NNSA under Contract No. 89233218CNA000001. It was supported by the Los Alamos National Laboratory (LANL) LDRD Program, and in part by the Center for Integrated Nanotechnologies, an Office of Science User Facility operated by the U.S. Department of Energy (DOE) Office of Science, in partnership with the LANL Institutional Computing Program for computational resources.  Additional computations were performed at the National Energy Research Scientific Computing Center (NERSC), a U.S. Department of Energy Office of Science User Facility located at Lawrence Berkeley National Laboratory, operated under Contract No. DE-AC02-05CH11231 using NERSC award ERCAP0020494.

\bibliographystyle{aps}

\providecommand{\noopsort}[1]{}\providecommand{\singleletter}[1]{#1}

\newpage\newpage
\begin{widetext}
\begin{Large}
\begin{center}
Supporting Information: Te Vacancy-Driven Anomalous Transport \\ in ZrTe$_{5}$ and HfTe$_{5}$
\end{center}
\end{Large}

\renewcommand{\thefigure}{S\arabic{figure}}
\renewcommand{\thetable}{S\arabic{table}}
\setcounter{figure}{0}
\setcounter{table}{0}

\begin{figure*}[h]
  \includegraphics[width=\textwidth]{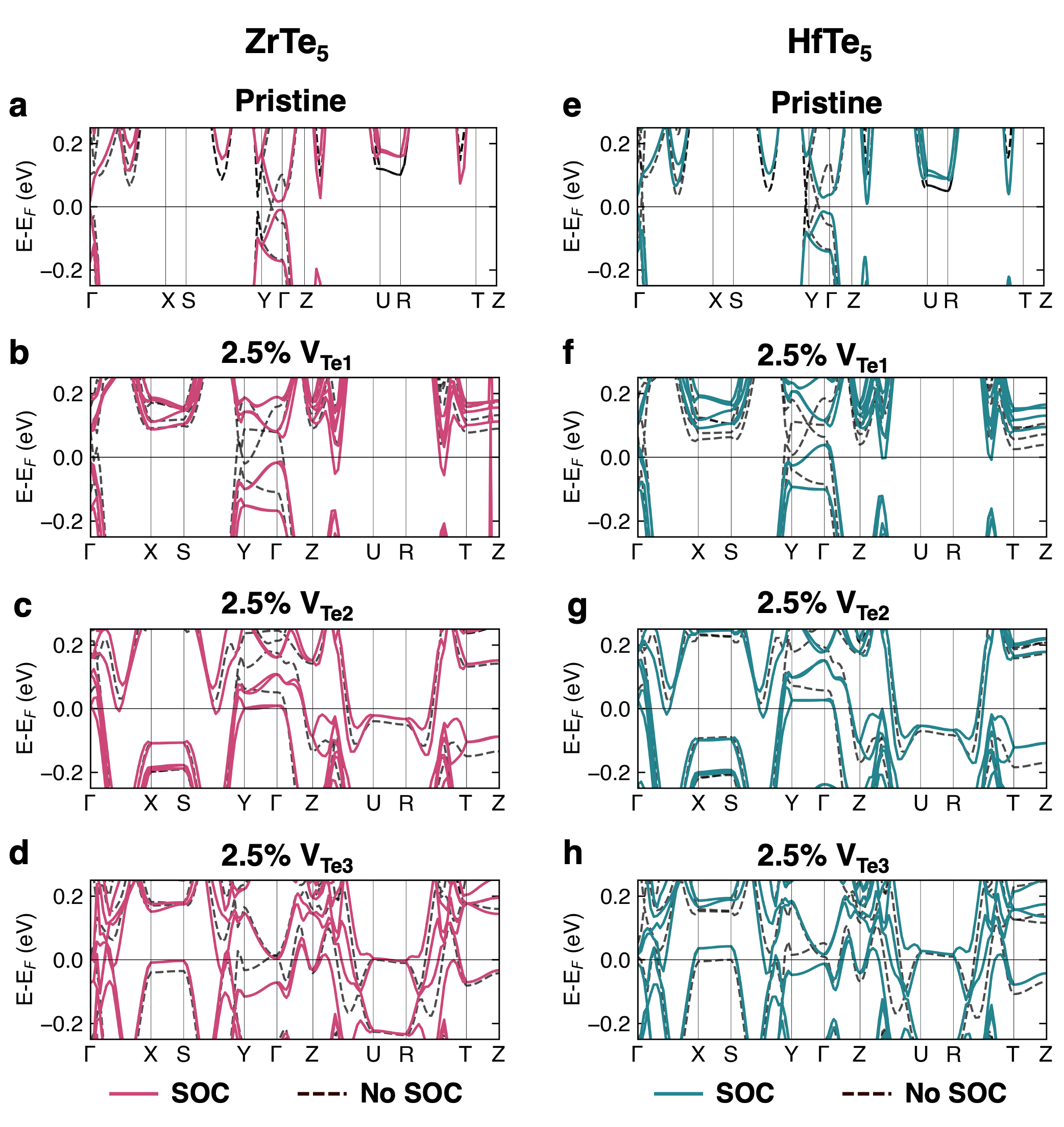}
  \caption{The DFT band structures near the Fermi energy of ZrTe$_{5}$(a-d) and HfTe$_{5}$(e-h) in the fully relaxed geometries for the pristine crystal structure (a,e) and the cases of a neutral Te vacancy in Te site 1 (b,f), Te site 2 (c,g), and Te site 3 (d,h). The band structures calculated both with (solid lines) and without (dashed lines) spin-orbit coupling (SOC) are plotted to illustrate how SOC affects the band structure around the Dirac point near the Fermi energy.}
  \label{fig:zrte5_hfte5_soc_no_soc_bs}
\end{figure*}

\begin{figure*}[h]
  \includegraphics[width=0.6\textwidth]{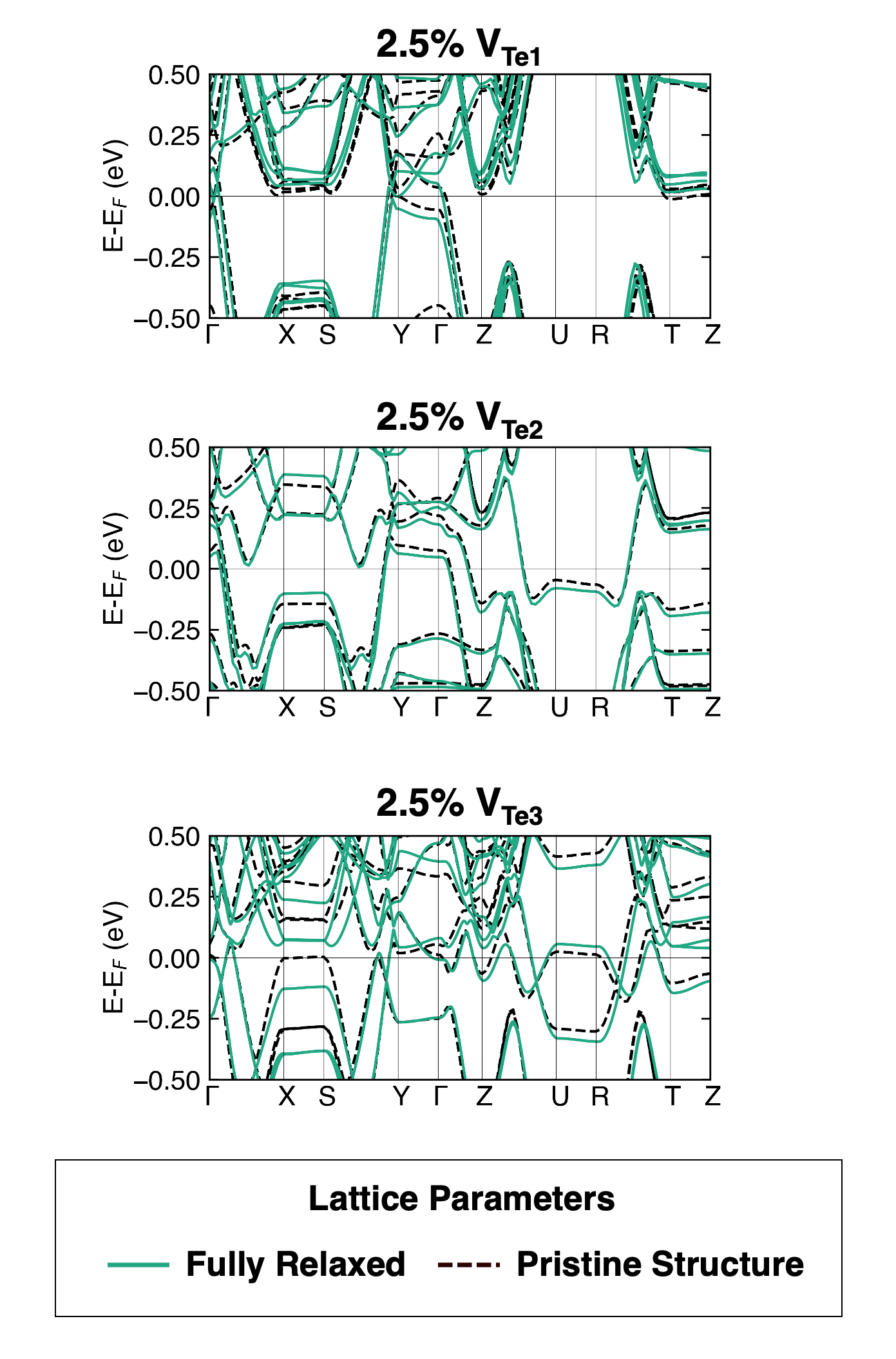}
  \caption{Band structure of HfTe$_{5}$ with Te vacancies calculated without spin-orbit coupling. The band structure calculated with fully relaxed lattice parameters for each type of Te vacancy is shown in green solid lines. The band structure calculated with the lattice parameters fixed at the lattice parameters of pristine HfTe$_{5}$ and only internal coordinates allowed to relax is shown in black dashed lines. In each case, there are bands in different parts of the Brillouin zone that shift to different energies but their shape and dispersion remain qualitatively similar. The corresponding lattice parameters are listed in \textbf{Table S1}.}
  \label{fig:hfte5_latt_param_bs}
\end{figure*}

\newpage
\newpage
\newpage
\newpage

\begin{table*}
 \caption{Lattice constants of a $2\,\times1\,\times1$ supercell of HfTe$_{5}$ for the pristine crystal structure and the crystal structures with each type of Te vacancy relaxed using DFT with Grimme-D3 dispersion corrections. The relative changes in the lattice parameters and volume are reproduced from Table 2 in the main text.}
 \begin{center}
  \begin{tabular}{c c c c c c c c}
    \hline
    & a (\textrm{\AA}) & b (\textrm{\AA}) & c (\textrm{\AA}) & $\Delta$a ($\%$) & $\Delta$b ($\%$) & $\Delta$c ($\%$) & $\Delta$V ($\%$) \\
    \hline
    Pristine & 8.000 & 14.711 & 13.596 & & & & \\
    V$_{\textrm{Te1}}$ & 7.990 & 14.720 & 13.594 & -0.12 & +0.06 & -0.02 & -0.08 \\
    V$_{\textrm{Te2}}$ & 7.913 & 14.707 & 13.555 & -1.09 & -0.03 & -0.31 & -1.42 \\
    V$_{\textrm{Te3}}$ & 7.978 & 14.727 & 13.469 & -0.27 & +0.11 & -0.93 & -1.10 \\
    \hline
  \end{tabular}
  \end{center}
  \label{table_latt}
\end{table*}

\begin{figure*}[h]
  \includegraphics[width=0.8\textwidth]{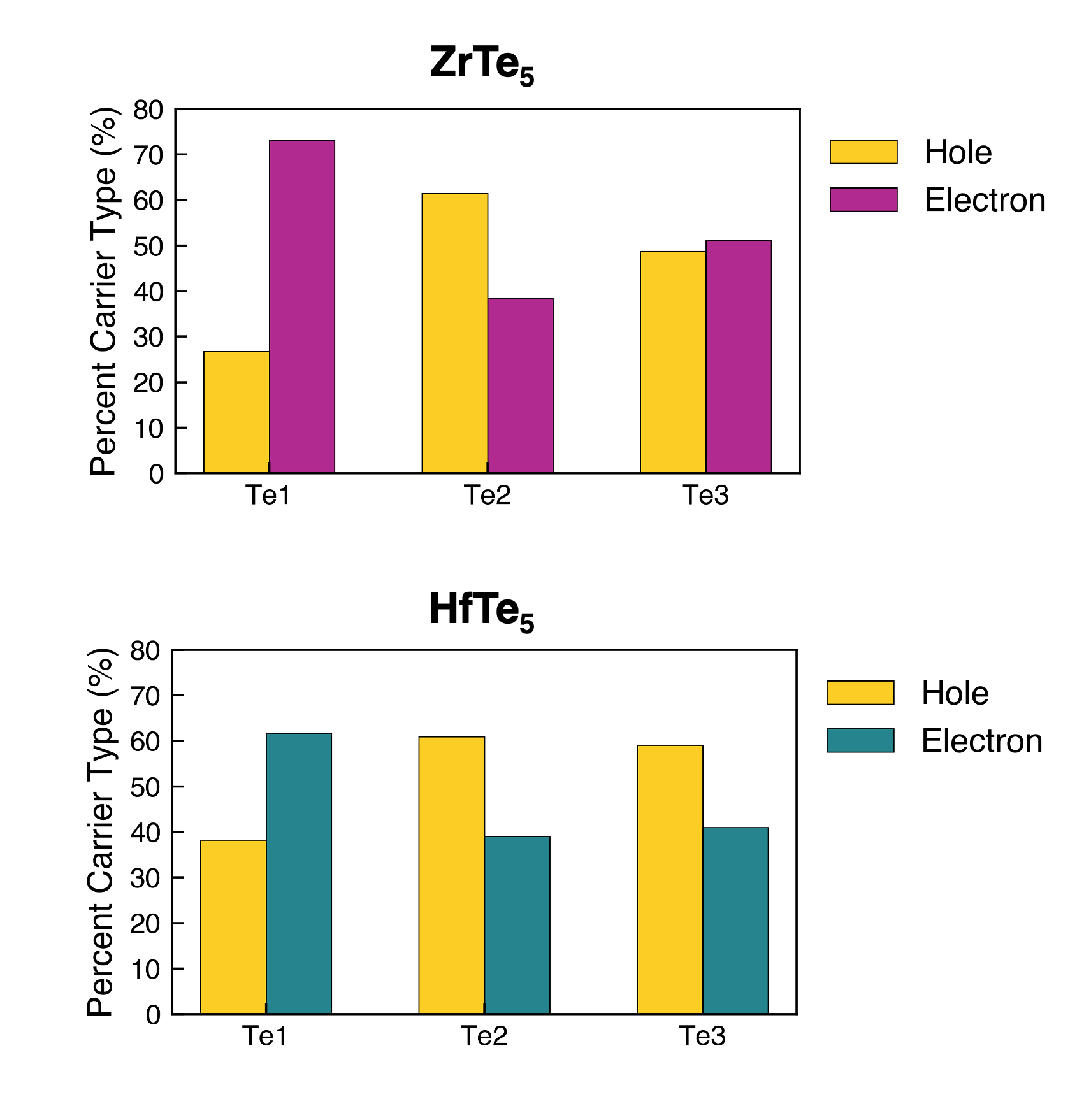}
  \caption{Percent of states near Fermi energy that correspond to hole-like or electron-like excess carriers for each type of Te vacancy site for ZrTe$_{5}$ (top) and HfTe$_{5}$ (bottom). The absolute number of excess carriers of each type is obtained by summing the number of Kohn-Sham states above the Fermi energy for bands that are below the Fermi energy in the pristine case and by summing the number of states below the Fermi energy for bands that were above the Fermi energy in the pristine case. Te vacancies on site 1, which is the most stable and expected to proliferate at the highest concentration, produce an excess of electron-like carrieres.}
  \label{fig:carrier_type}
\end{figure*}

\end{widetext}

\end{document}